\documentclass[bibyear]{aa} 
\usepackage[varg]{txfonts}

\usepackage{graphicx}
\usepackage{multirow}
\usepackage{colortbl}
\usepackage[table,x11names]{xcolor}
\usepackage{natbib}
\usepackage{txfonts}
\DeclareMathOperator{\sech}{sech}
\usepackage{threeparttable}
\usepackage{amstext}

\begin{document}

   \title{Gravitational potential energy of a multi-component galactic disk}
      \titlerunning{Potential energy of a multi-component disk}
\authorrunning{Suchira Sarkar \& Chanda J. Jog}

   \author{Suchira Sarkar\inst{1}\fnmsep\thanks{suchira@iisc.ac.in}
          \and
           Chanda J. Jog\inst{1}}

   \institute { Department of Physics,
    Indian Institute of Science, Bangalore 560012, India}
    
\date{Received xxxx; accepted yyyy}

\abstract{
We calculate ab initio the gravitational potential energy per unit area for a gravitationally coupled multi-component galactic disk of stars and gas, which is given as the integration over vertical density distribution, vertical gravitational force, and vertical distance. This is based on the method proposed by Camm for a single-component disk, which we extend here for a multi-component disk by deriving the expression of the energy explicitly at any galactocentric radius $R$. For a self-consistent distribution, the density and force are obtained by jointly solving the equation of vertical hydrostatic equilibrium and the Poisson equation. Substituting the numerical values for the density distribution and force obtained for the coupled system, in the derived expression of the energy, we find that the energy of each component remains unchanged compared to the energy for the corresponding single-component case. We explain this surprising result 
by simplifying the above expression for the energy of a component analytically, which turns out to be equal to the surface density times the squared vertical velocity dispersion of the component. However, the energy required to raise a unit test mass to a certain height $z$ from the mid-plane is higher in the coupled case. The system is therefore more tightly bound closer to the mid-plane, and hence it is harder to disturb it due to an external tidal encounter.}


\maketitle

\section{Introduction}           
\label{sec:chap4_intro}
The vertical structure of the stellar disk in a galaxy has been studied as a self-gravitating, single-component, isothermal disk in the literature, where its self-consistent vertical density distribution is given by a $\sech^{2}$ form (\citealt{1942_Spitzer}). However, a real galactic disk is a multi-component system of gravitationally coupled stars and interstellar gas (HI and $\mathrm{H_{2}}$) embedded in the potential of the dark matter halo.
The self-consistent vertical distribution of stars in such a system is determined by the joint gravitational potential of stars, gas, and the dark matter halo (\citealt{2002_Narayan_Jog}, \citealt{2018_Sarkar}), instead of its self-gravitational force alone. The joint potential of the system is found to constrain the distribution of stars towards the mid-plane, and thus increases the mid-plane density value and decreases the vertical disk thickness (\citealt{2018_Sarkar}). In the inner Galaxy, gas plays the dominant role in constraining the distribution of stars (\citealt{2018_Sarkar}). Although gas contains 10–15\% of the disk mass (\citealt{1991_Young_Scoville}; \citealt{1998_Binney_Merrifield}), it forms a thin layer about the mid-plane due to its low vertical velocity dispersion and therefore can affect the vertical distribution of stars significantly in the inner Galaxy. Through gravitational coupling, stars also constrain the gas distribution in a similar way. Stars, being a much more massive component, have a stronger effect on gas. 
 Therefore we expect the vertical distribution of stars and gas to be more strongly bound in a coupled system than in single-component cases, and thus it is more robust against perturbations or external tidal interactions.

In this context, it is interesting to study the gravitational potential energy of coupled stellar and gas distribution. 
The higher the potential energy of the vertical distribution of stars (gas), the more difficult it should be to distort the disk in external tidal encounters. Motivated by the results of the constraining effect of the joint potential, we expect the potential energy of a component to be higher in the gravitationally coupled multi-component system than in its single-component case. 
With this aim, we study the potential energy per unit area of the vertical distribution of stars, and gas in the single-component self-gravitating cases as well as in the coupled, multi-component system. We consider a gravitationally coupled stars-plus-gas disk, a two-component system, in the inner Galaxy, and explicitly derive the expression for the potential energy per unit area of the disk, following the method proposed by \citet{1967_Camm} for the single-component case. We study how the energy corresponding to each component is affected by the gravitational coupling between them. Surprisingly, we find that the energy of each component remains unchanged, and the physical reason for this is explained in the paper. However, despite this, we found that stars and gas are more strongly bound to the mid-plane in the coupled case, and thus are less susceptible to external tidal distortions. For completeness, we have also studied the potential energy per unit area of the disk for a three-component case, that is, for a gravitationally coupled disk of stars and two gas components.

In the outer Galaxy, on the other hand, the dark matter halo plays the dominant role in constraining the vertical distribution of stars and gas significantly (\citealt{2018_Sarkar}). For simplicity, we did not include the dark matter halo to study the multi-component system here because the dark matter halo is shown to have a less significant effect than gas (stars) on the vertical distribution of stars (gas) in the inner Galaxy (\citealt{2002_Narayan_Jog}, \citealt{2018_Sarkar}). 

We also note that we ignored the bulge in the inner Galaxy. Our theoretical model is applied from a galactocentric radius of 4 kpc onward. The bulge is not a dominant gravitating component in the region studied here (\citealt{ghosh_2016}; \citealt{1995_Blum}).

We show the formulation in Section \ref{sec:chap4_formulate}, results in Section \ref{sec:chap4_results}, and give discussion and conclusions in Sections \ref{sec:chap4_discussions} and \ref{sec:chap4_conclusions}, respectively.
The formulation of the potential energy, given in Section \ref{sec:chap4_formulate}, is general and applicable to any two (three)-component disk, for instance, for an n-component stellar disk, even though we apply it to the specific and observationally motivated case of a stars-plus-gas disk. In other words, the formulation of the energy does not in any way involve the specific physical nature and properties of gas, such as dissipation or low dispersion.

\section{Formulation of the problem}
\label{sec:chap4_formulate}
\subsection{Gravitational potential energy of a single-component isothermal galactic disk}
\label{sec:formulate_chap4_single}

First we discuss the formulation of the gravitational potential energy of a single component self-gravitating galactic disk that can be taken to be a stars-alone or a gas-alone disk. For a mass distribution contained in a finite volume of space, the gravitational force decreases as $\sim 1/r^{2}$ at a large distance and hence the work done to bring a unit mass from infinity to a certain finite distance is obtained to be a finite quantity. The gravitational potential energy of the mass distribution in such a case is defined as the energy released in assembling the finite system from an infinitely dispersed state. For a galactic disk, however, the mass distribution is stratified in plane parallel layers and is infinite on the x-y plane. The vertical gravitational force for such a stratified mass distribution remains constant at a value of $-2\pi G \Sigma$ at large $z$ at any galactocentric radius $R$, $\Sigma$ being the surface density of the disk at that $R$, beyond the vertical extent of the mass distribution (and within the disk approximation limit). This results in the energy released in bringing a unit mass from infinity to a certain finite height to be infinite. Therefore, instead, the state of complete collapse of the disk mass on the $z=0$ plane is defined here to be the state of the zero potential energy, and the work required to build the disk from that state is considered to be the potential energy stored in the disk. For a detailed discussion of this point and the derivation of Eq. (\ref{chap4_eq:1}) (see below) for a single-component case, see \citet{1967_Camm}.

We note that the mass contained in a column of unit cross section perpendicular to the mid-plane is finite. Therefore the potential energy of the disk is defined in terms of the energy contained in a column of unit cross section, that is, as the potential energy per unit area of the disk.
We use the galactocentric cylindrical coordinates ($R, \phi, z$) and consider the disk to be axisymmetric. 
 
The mathematical expression for the gravitational potential energy per unit area of a stars-alone disk has been derived in \citet{1967_Camm} as

\begin{equation}
W=-\int_{-\infty}^{+\infty} \rho(z)\frac{d \Phi}{d z}z \ dz \label{chap4_eq:1}
,\end{equation}

\noindent where $\rho(z)$ is the vertical mass density distribution of stars, $\Phi$ is the gravitational potential of the disk, and $d\Phi/dz$ is taken to be the force per unit mass ($K_{z}$) due to the self-gravity of the stellar disk, acting along the negative $z$ direction. This represents the self-gravitational energy per unit area of the disk. A similar expression was used by \citet{2010_Pestana} and \citet{1984_Fridman} to calculate the gravitational potential energy of a single-component disk, but the expression was not derived.

The above expression in \citet{1967_Camm} was obtained by using the Poisson equation in the form of $d^{2}\Phi/dz^{2}=-4\pi G \rho(z)$, where the vertical force was defined as $K_{z}=d \Phi/dz$, which is a negative quantity. However, we adopted the standard notation here that is routinely used in the literature, where the Poisson equation is given by $d^{2}\Phi/dz^{2}=4\pi G \rho(z)$ and the vertical force is defined by $K_{z}=-d\Phi/dz,$ where $d\Phi/dz$ is positive. Following the treatment in \citet{1967_Camm}, we therefore derive the expression of the energy as 

\begin{align}
W &= \int_{-\infty}^{+\infty} \rho(z)\frac{d\Phi}{d z}z \ dz \nonumber \\ 
&= -\int_{-\infty}^{+\infty} \rho(z)K_{z}z \ dz. \label{chap4_eq:2}
\end{align}

\noindent A detailed derivation for a multi-component case is given in Section \ref{sec:chap4_multi}. The negative sign in front of Eq.\ref{chap4_eq:1} or Eq.\ref{chap4_eq:2} indicates that the energy is  positive. 

We note a few important points here. While deriving Eq.\ref{chap4_eq:1}, \citet{1967_Camm} used only the z-term in the Poisson equation.  
We show that this treatment is justified for a thin galactic disk. We show that the z-term of the Poisson equation is much greater than the R-term \footnote{$\mathrm{\text{The }R~term}$ = $\mathrm{(1/R)(2V_{c})dV_{c}/dR} = \mathrm{2(B^{2} - A^{2})}$ (see e.g.,\citet{1968_Mihalas_Routly}), where A and B are Oort’s constants, and $\mathrm{V_{c}}$ is the rotational velocity. The z term is ($4\pi G \rho_{0} - R~term $), where  $\rho_{0}$ is the mid-plane (z=0) density, obtained as $\mathrm{\rho_{0}= \Sigma}/(2z_{d})$. Here $\Sigma$ is the surface density of a radially exponential stellar disk, and $z_{d}$ is the scale height. The values of A, B, and $z_{d}$ in the solar neighbourhood (R= 8.5 kpc) on the mid-plane are taken from \citet{2008_BT} (Table 1.2), and $\Sigma$ is calculated as in Section \ref{sec:chap4_input} in the paper. The R-term and z-term are found to be -130.56 $\mathrm{km^{2}/s^{2}/kpc^{2}}$ and 4204.6 $\mathrm{km^{2}/s^{2}/kpc^{2}}$ , respectively. Thus the R-term is 3.1\% of the z term and hence is negligible.  A similarly low ratio of R to z term holds true at other radii, e.g., beyond R=4 kpc and up to R=10 kpc. We checked this using the observed values of $\mathrm{V_{c}}$, $\mathrm{dV_{c}/dR}$ from \citet{Eilers_2019}, and $\rho_{0}$ obtained in the similar way assuming a constant $z_{d}$ with a value as at R=8.5 kpc.}. Thus the density distribution, force, and energy become only z-dependent quantities. 
We followed the same approach while deriving the energy for the multi-component case in Section \ref{sec:chap4_multi}. We also note that \citet{1967_Camm} implicitly assumed the disk to have a constant radial surface density, whereas we considered realistic stellar and gas disks of radially varying surface density. This does not affect the derivation of the energy or the application of the model because the calculation is local. We used the surface density value at any given radius as a  local constraint to obtain $\rho(z)$ at a given R, which was then used in the expression for the energy, as discussed below and in Section \ref{sec:chap4_theory}. Thus the calculation is local, and the energy is independent of the value of surface density at other radii.

We note that although Eq.\ref{chap4_eq:2} is derived in \citet{1967_Camm}, using explicitly only the Poisson equation, for a disk in vertical hydrostatic equilibrium, $\rho(z)$ and $K_{z}$ are related to each other and have to be obtained by solving the joint hydrostatic balance-Poisson equation. These solutions are required to obtain a numerical value for $W$. We show this set of equations below for a single-component self-gravitating  isothermal disk.

 We assume the vertical velocity dispersion ($\sigma_{z}$) of the component to be isothermal along $z$. The vertical hydrostatic balance equation for a single-component isothermal self-gravitating disk is given by 

\begin{equation}
\frac{\sigma^{2}_{z}}{\rho}\frac{d\rho}{dz}=K_{z}\label{chap4_eq:3}
.\end{equation}

\noindent 
The Poisson equation for a single-component galactic disk is given as

\begin{equation}
\frac{d^{2}\Phi}{dz^{2}}=4\pi G \rho(z)\label{chap4_eq:4}
.\end{equation}

\noindent We combine these two equations to obtain the joint hydrostatic balance-Poisson equation 

\begin{equation}
\sigma^{2}_{z}\frac{d}{dz}\left[\frac{1}{\rho}\frac{d\rho}{dz}\right] = -4 \pi G \rho\label{chap4_eq:5}
.\end{equation}
\noindent The analytical solution of this equation in form of $\sech^{2}(z/z_{0})$ was obtained by \citet{1942_Spitzer}. The analytical expressions of the density distribution and the force are given as 

\begin{equation}
\rho(z)=\rho_{0}\sech^{2}(z/z_{0}); |K_{z}|=2\frac{\sigma^{2}_{z}}{z_{0}}\tanh\left(\frac{z}{z_{0}}\right), \label{chap4_eq:6}
\end{equation}

\noindent respectively, where $z_{0} = \left({\sigma^{2}_{z}}/2\pi G \rho_{0}\right)^{1/2}$.
Using the analytical form of the solution $\rho(z)$ versus $z$, we calculate $\rho_{0}$ (mid-plane density) and $z_{0}$ (scale-height) by integrating $\rho(z)$ versus $z $, using the constraint of the observed surface density, defined as $\Sigma=\int_{-\infty}^{+\infty}\rho\ dz$, at a given galactocentric radius R. The extent of $z$ in the numerical integration is chosen such that the solution obtained is saturated.
The obtained value of $z_{0}$ in turn is used to calculate $|K_{z}|$ up to the same limit of $z$. Thus using $\rho(z)$ and $K_{z}$ we calculate $W$ using Eq. \ref{chap4_eq:2}, which gives us the saturated value of the energy.

However, for a gravitationally coupled two-component system, the joint hydrostatic balance-Poisson equation has to be solved numerically to obtain $\rho(z) ~\text{versus }z $ and $|K_{z}| ~\text{versus } z$. This is discussed in Section \ref{sec:chap4_distribution}.

\subsection{Gravitational potential energy of a multi-component gravitationally coupled isothermal galactic disk}
\label{sec:chap4_multi}
In this section, we explicitly derive the expression of the gravitational potential energy per unit area for a realistic model of galactic disk that is a multi-component system of gravitationally coupled stars and gas, taken at a given galactocentric radius $R$. We consider the disks of stars and gas to be coplanar with the same mid-plane at z=0. We consider the disk to be a thin disk and therefore consider only the z-term in the corresponding Poisson equation for the multi-component system (see Section \ref{sec:formulate_chap4_single}). We show the detailed derivation for a two-component system, consisting of stars and one gas component, and discuss the same for a three-component system of stars and two gas components in Section \ref{sec:chap4_three}. The formulation in each case is done following the same steps as in the treatment for the stars-alone case in \citet{1967_Camm}. In the following, the subscripts $i=s,g$ in the quantities $\rho(z)$, $\Sigma$, $\Phi$ denote stars and gas, respectively. 

We consider the galactic disk to be a gravitationally coupled stars-plus-gas system where the vertical distribution of each component is determined by the joint gravitational force from the stars and gas. We take the vertical velocity dispersion of each of the two components
to be isothermal. We assume that initially, all the mass of the stars-plus-gas system lies on the $z=0$ plane, and the potential energy $W$ of this system is assumed to be zero based on the same arguments as discussed in Section \ref{sec:formulate_chap4_single}. Now the work done per unit area to build a gravitationally coupled stars-plus-gas disk together, at the same time, is stored as the gravitational potential energy per unit area of the stars-plus-gas disk.

Below we present the detailed derivation of the energy per unit area of this coupled stars-plus- gas disk following all the steps used for the stars-alone case in \citet{1967_Camm}. As stated earlier in Section \ref{sec:formulate_chap4_single}, we used $d^{2}\Phi/dz^{2}=4\pi G \rho(z)$ and $K_{z}=-d\Phi/dz$ in each step of the derivation, as shown below. 

For a gravitationally coupled stars-plus-gas disk, the Poisson equation is given as 

\begin{align}
\frac{d^{2}\Phi_{s}}{dz^{2}}+\frac{d^{2}\Phi_{g}}{dz^{2}} &= 4\pi G (\rho_{s}+\rho_{g}) \nonumber \\ 
Or, \frac{d^{2}\Phi_{\mathrm{coupled}}}{dz^{2}} &= 4\pi G (\rho_{s}+\rho_{g}). \label{chap4_eq:7}
\end{align}

In the first step, we integrate Eq. (\ref{chap4_eq:7}) and derive an expression for the vertical force of the coupled system to be (after some algebraic manipulations) 

\begin{equation}
\frac{d\Phi_{\mathrm{coupled}}}{dz} = 2\pi G\int_{-\infty}^{z} (\rho_{s}+\rho_{g}) dz - 2\pi G\int_{z}^{\infty} (\rho_{s}+\rho_{g}) dz \label{chap4_eq:8}
\end{equation}

\begin{align}
K_{z,\mathrm{coupled}} &= -\frac{d\Phi_{\mathrm{coupled}}}{dz} \nonumber \\
&= 2\pi G\int_{z}^{\infty} (\rho_{s}+\rho_{g}) dz - 2\pi G\int_{-\infty}^{z} (\rho_{s}+\rho_{g}) dz. \label{chap4_eq:9}
\end{align} 

\noindent Now while building up the gravitationally coupled stars-plus-gas disk from $z=0$, at any intermediate step, only a fraction of the total mass of stars-plus-gas is distributed along $z$, denoted by the density distribution of $\epsilon(\rho_{s}+\rho_{g})$ in the region $z>0$, where $\epsilon$ lies between 0 and 1. The rest of the mass still lies on the $z=0$ plane with the mass per unit area as $(\Sigma_{s}+\Sigma_{g})-\epsilon\int_{0}^{\infty}(\rho_{s}+\rho_{g}) dz$. The value of $\epsilon$ at any intermediate step has no specific physical meaning. It was used only to denote a fraction of the sum of the final vertical distribution of stars-plus-gas (i.e. $(\rho_{s}+\rho_{g})$), at an intermediate step. We also note that $\epsilon$ has been taken to be independent of $z$ in \citet{1967_Camm}. 

At any intermediate step, the force against which the work is being done is due to the joint stars-plus-gas system at that step. For any height $x(>0)$ above the $z=0$ plane, the force ($K_{\mathrm{fraction,coupled}}$), for such a system can be written as

\medskip

\begin{align*}
K_{\mathrm{fraction,coupled}} &= 2\pi G\int_{x}^{\infty} \epsilon(\rho_{s}+\rho_{g}) dz - 2\pi G\int_{0}^{x} \epsilon(\rho_{s}+\rho_{g}) dz \\
&-2\pi G\bigg((\Sigma_{s}+\Sigma_{g})- \int_{0}^{\infty} \epsilon(\rho_{s}+\rho_{g}) dz \bigg).  \nonumber
\end{align*}
\noindent Here the first two terms account for the force from the stars-plus-gas distribution along $z>0,$ and the rest of the terms account for the force due to the rest of the mass lying on the $z=0$ plane. Now, we can write
\begin{align*}
K_{\mathrm{fraction,coupled}} &=2\pi G\int_{x}^{\infty} \epsilon(\rho_{s}+\rho_{g}) dz-2\pi G\int_{0}^{x} \epsilon(\rho_{s}+\rho_{g}) dz \\
&-2\pi G\bigg((\Sigma_{s}+\Sigma_{g}) -\int_{0}^{\infty} \epsilon(\rho_{s}+\rho_{g}) dz \bigg)  \\
&-2\pi G\int_{-\infty}^{0} \epsilon(\rho_{s}+\rho_{g}) dz +2\pi G\int_{-\infty}^{0} \epsilon(\rho_{s}+\rho_{g}) dz \\
&= 2\pi G\int_{x}^{\infty} \epsilon(\rho_{s}+\rho_{g}) dz - 2\pi G\int_{-\infty}^{x} \epsilon(\rho_{s}+\rho_{g}) dz \\
&-2\pi G(\Sigma_{s}+\Sigma_{g})+2\pi\epsilon G(\Sigma_{s}+\Sigma_{g}) \\
&= 2\pi G\int_{x}^{\infty} \epsilon(\rho_{s}+\rho_{g}) dz - 2\pi G\int_{-\infty}^{x} \epsilon(\rho_{s}+\rho_{g}) dz \\
&-2\pi G(\Sigma_{s}+\Sigma_{g})(1-\epsilon). 
\end{align*}    

\noindent The above equation can be rewritten, using Eq.(\ref{chap4_eq:8}), as 

\begin{equation}
K_{\mathrm{fraction,coupled}}=-\epsilon\frac{d \Phi_{\mathrm{coupled}}}{dx}-2\pi G (\Sigma_{s}+\Sigma_{g})(1-\epsilon). \label{chap4_eq:10}
\end{equation}

\noindent Now the work done per unit area of the z-plane to raise the density of stars plus gas between $z$ and $z+\delta z$ from $\epsilon(\rho_{s}+\rho_{g})$ to $(\epsilon + \delta \epsilon)(\rho_{s}+\rho_{g})$ is the work done in raising the mass $(\rho_{s}+\rho_{g}) \delta\epsilon  \delta z$ from $z=0$ to the chosen $z$ plane, and is given by

\begin{equation}
-(\rho_{s}+\rho_{g}) \delta \epsilon \delta z \int_{x=0}^{z} - \bigg\{\epsilon\frac{d \Phi_{\mathrm{coupled}}}{dx}+2\pi G (\Sigma_{s}+\Sigma_{g})(1-\epsilon)\bigg\} \,dx. \label{chap4_eq:11}
\end{equation}

\noindent We now calculate the potential energy per unit area of the two-component disk of $\rho_{s}+\rho_{g}$ by integrating the above equation over $\epsilon$ (from 0 to 1) and $z$ (from $-\infty$ to $\infty$), following the steps shown in \citet{1967_Camm} for the one-component case.

\noindent We obtain the expression for the energy as
\begin{equation}
W_{\mathrm{coupled}} = \int_{-\infty}^{\infty}z~ \frac{d\mathrm{\Phi_{coupled}}}{dz}(\rho_{s}+\rho_{g})  \ dz.  \label{chap4_eq:12}
\end{equation}

\noindent \textup{}This is the most important result of this paper. Now the above equation can be further expressed as 
\begin{align}
W_{\mathrm{coupled}} &= \int_{-\infty}^{\infty}z\frac{d \Phi_{\mathrm{coupled}}}{d z} \rho_{s} \ dz +\int_{-\infty}^{\infty}z\frac{d \Phi_{\mathrm{coupled}}}{d z} \rho_{g} \ dz   \nonumber \\
&= -\int_{-\infty}^{\infty}z K_{z,\mathrm{coupled}} ~\rho_{s} \ dz - \int_{-\infty}^{\infty}zK_{z,\mathrm{coupled}} ~\rho_{g} \ dz.  \nonumber 
\end{align}
Using the symmetry of $\rho(z) vs.z$ about $z=0$, this can be written as

\begin{equation}
W_{\mathrm{coupled}} = -2\int_{0}^{\infty}z K_{z,\mathrm{coupled}} ~\rho_{s} \ dz - 2\int_{0}^{\infty}zK_{z,\mathrm{coupled}} ~\rho_{g} \ dz \label{chap4_eq:13}  
,\end{equation}

\noindent where the first integration can be considered to represent the potential energy per unit area of the stellar disk in the coupled stars-plus-gas system, and the second integration to represent the potential energy per unit area of the gas disk in the coupled stars-plus-gas system. In the limit of $\rho_{g} \to 0$ or $\rho_{s} \to 0$, this goes over to the one-component case (see Eq.\ref{chap4_eq:2} and \ref{chap4_eq:9}). 
The separation of energy into the two components, as can be seen from the above equation, may appear somewhat surprising, but we note that all the expressions starting from Eq.(\ref{chap4_eq:9}) can be written in separable form for stars and gas, except that the components are being built against the same coupled force. 

The above formulation can be used for any $n$ -component system (e.g. for n>2). For illustration, we show this for a three-component disk consisting of stars and two gas components in a similar fashion as discussed in Section \ref{sec:chap4_three}.

\subsection{Theoretical model for the vertical distribution for a  multi-component disk and input parameters}
\label{sec:chap4_theory}

\subsubsection{Self-consistent vertical distribution in a multi-component disk}
\label{sec:chap4_distribution}

For a gravitationally coupled two-component stars-plus-gas disk, the hydrostatic balance of each of the components is determined by the joint gravitational force of stars and gas, and is given by

\begin{equation}
\frac{\sigma^{2}_{z,i}}{\rho_{i}}\frac{d\rho_{i}}{dz}=K_{z,s}+K_{z,g}\label{chap4_eq:14}
,\end{equation}
\noindent where $i$ represents stars $(s)$ or gas $(g)$ and the right-hand side of the equation represents the vertical force from the coupled stars-plus-gas case. 
For a gravitationally coupled stars-plus-gas galactic disk, the Poisson equation is given as 

\begin{align}
\frac{d^{2}\Phi_{s}}{dz^{2}}+\frac{d^{2}\Phi_{g}}{dz^{2}} &= 4\pi G (\rho_{s}+\rho_{g}) \nonumber \\
Or, \frac{d^{2}\Phi_{\mathrm{coupled}}}{dz^{2}} &= 4\pi G (\rho_{s}+\rho_{g}). \label{chap4_eq:15}
\end{align}

\noindent We combine these two equations to write the joint hydrostatic balance-Poisson equation as

\begin{equation}
\frac{\mathrm{d}^{2}\rho_{i}}{\mathrm{d}z^{2}} = \frac{\rho_{i}}{\sigma^{2}_{z,i}}\left[-4\pi G\left(\rho_{\mathrm{s}}+ \rho_{\mathrm{g}}\right)\right] +\frac{1}{\rho_{i}}\left(\frac{\mathrm{d}\rho_{i}}{\mathrm{d}z}\right)^{2}  \label{chap4_eq:16}
.\end{equation}

\noindent These coupled equations are solved numerically using the fourth-order Runge-Kutta method to obtain $\rho_{i}(z) vs.z $, simultaneously for each $i$th component in an iterative fashion, as discussed in \citet{2002_Narayan_Jog} and \citet{2018_Sarkar}, until the fifth decimal convergence in the solutions. To solve the equations at a radius, we use the observed surface density of each component as one boundary condition, and $d\rho_{i}/dz=0$ at $z=0$ as the other boundary condition, where the latter is true for any realistic distribution that is homogeneous very close to the mid-plane. The vertical distributions for a three-component system can be obtained following a similar method, as discussed in Section \ref{sec:chap4_three}.

\subsubsection{Input parameters}
\label{sec:chap4_input}

The formulation presented so far is general. Here we apply it for the Milky Way. We considered HI as the gas component here to study the two-component system of the stars-plus-gas disk, and chose the solar radius, taken to be at $R$=8.5kpc, to illustrate the results. The stellar disk was taken to be exponential with the central surface density $\Sigma_{0}=640.9M_{\odot}pc^{-2}$ and radial scale length $R_{D}=3.2kpc$ \citep{1998_Mera}. Thus the surface density value of the stellar disk is 45.0 $M_{\odot}pc^{-2}$ at the solar radius.
 
The radial velocity dispersion values of stars on the mid-plane were obtained observationally by \citet{1989_Lewis_Freeman} up to $R$=16kpc. It falls off exponentially with radius as $\sigma_{R,s}=105\exp(-R/8.7 kpc) kms^{-1}$. We calculated the corresponding vertical velocity dispersion value ($\sigma_{z,s}$) on the mid-plane by assuming the vertical to radial dispersion ratio to be 0.45 (\citealt{1998_Dehnen_Binney}; \citealt{2000_Mignard}), as observed in the solar neighbourhood. Thus $\sigma_{z,s}$ at the solar radius is calculated to be 17.8$kms^{-1}$. The dispersion was taken to be isothermal along $z$.
 
The surface density value of the HI disk was taken to be 5.5 $M_{\odot}pc^{-2}$ (\citealt{1987_Scoville}). The vertical velocity dispersion of HI at the solar radius was taken to be 8 $kms^{-1}$ and isothermal, based on the values given by \citet{1978_Spitzer} for the Galaxy, and \citet{1984_Lewis} for nearly 200 face-on galaxies. 
The corresponding input parameters to study a three-component system are discussed in Section \ref{sec:chap4_three}.

\section{Results}
We first show the results for the two-component stars-plus-gas (HI) disk in Sections \ref{sec:chap4_result_multi},  \ref{sec:analytical}, and  \ref{sec:chap4_unit_mass} in detail. We then show the results for a three-component disk for completeness, taking stars and two gas components (HI and $\mathrm{H_{2}}$) in Section \ref{sec:chap4_three}.
\label{sec:chap4_results}

\subsection{Calculation of the potential energy of the two-component stars-plus-gas disk}
\label{sec:chap4_result_multi}
\begin{figure*} 
\centering
\includegraphics[height=2.3in,width=3.2in]{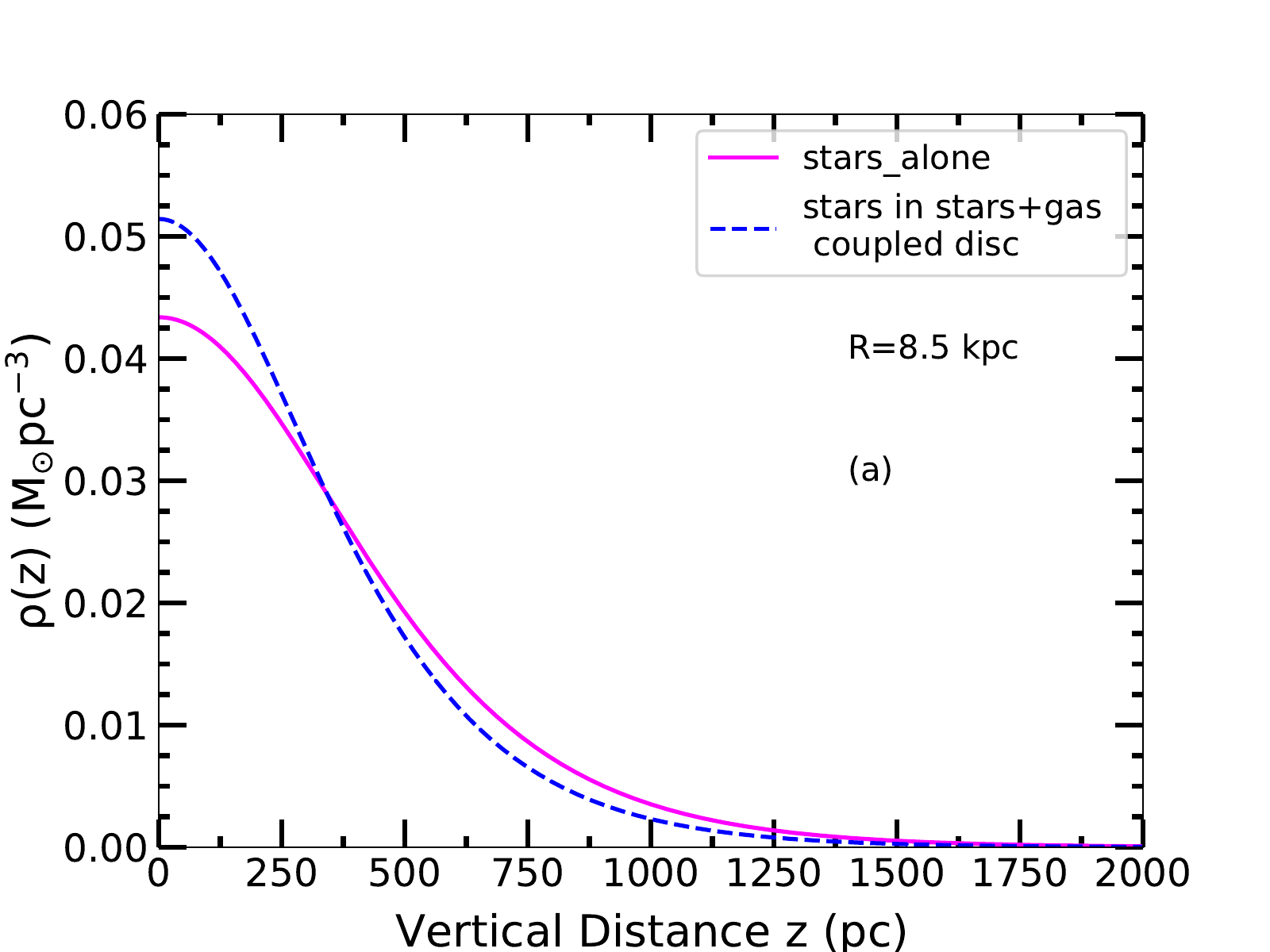}
\medskip
\includegraphics[height=2.3in,width=3.2in]{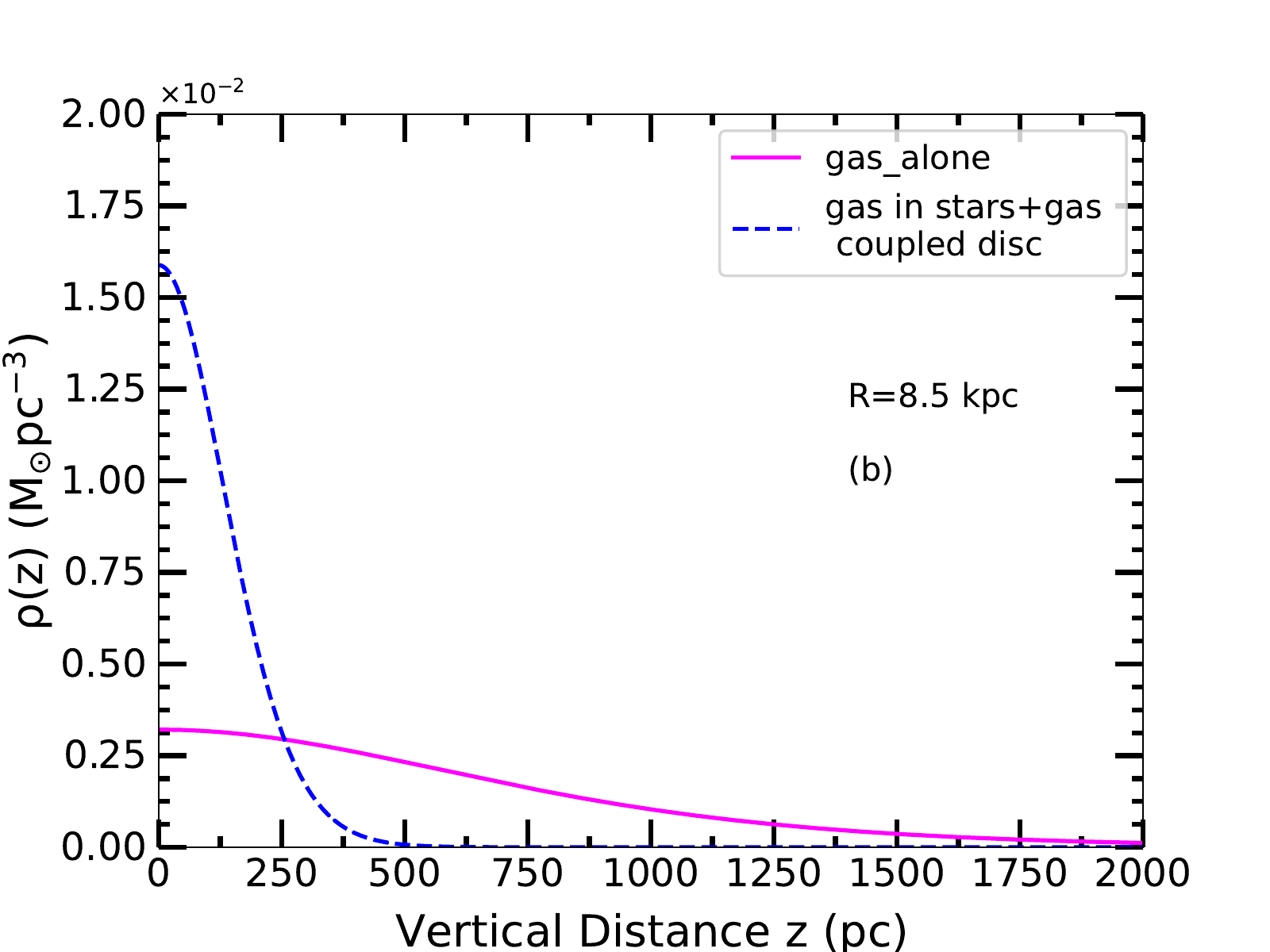}
\bigskip
\caption{Stellar and gas (HI) vertical density distribution, $\rho(z)$ vs. $z$ at $R$= 8.5 kpc (panels a and b, respectively). In each plot, the solid curve represents the vertical distribution in the one-component case. The dashed curve represents the distribution obtained in the gravitationally coupled stars-plus-gas system. The coupled gravitational force of the star- plus-gas system constrains each distribution towards the mid-plane, i.e. increases the mid-plane density value of the distribution, which causes the scale height to become smaller and the curve to become steeper.}
\label{chap4_label1}
\end{figure*}

We calculated the vertical distributions of stars and gas (HI) at $R$= 8.5 kpc first, taking each as a single-component self-gravitating system (as discussed in Section \ref{sec:formulate_chap4_single}), and then for the coupled two-component system of stars plus gas (as discussed in Section \ref{sec:chap4_theory}). We compare the vertical density distributions of stars in Fig.\ref{chap4_label1}(a) and gas in Fig.\ref{chap4_label1}(b) in these two cases.
We note that the vertical distribution of each of the components solved in the coupled system is constrained towards the mid-plane. It has a higher mid-plane density that falls off more sharply along $z$ and therefore has a smaller disk thickness than those in the corresponding one-component case. This is due to an additional gravitational force from a second component in the coupled system. We show the mid-plane density (up to third decimal place) and the half width at half maximum (HWHM) values of the density distribution that define the disk thickness in Table \ref{chap4_table:1}.
We note that due to the higher mass content of the stellar distribution, stars affect gas more strongly.

\begin{table} 
\begin{threeparttable}
\caption{Results for mid-plane density values (up to the third decimal place) and HWHM for stellar and gas distributions in their single-component vs. coupled stars-plus-gas cases} 
\label{chap4_table:1}
\centering
  \begin{tabular}{l l l l }      
\hline \hline
$\mathrm{\rho_{0}(M_{\odot}\mathrm{pc^{-3}})}$ \\  
\hline \hline
$\mathrm{stars-alone}$ & $\mathrm{stars-coupled}$ & $\mathrm{gas-alone}$ & $\mathrm{gas-coupled}$ \\
\hline                                                                                                                                   
0.043   &       0.051   &       0.003   & 0.016 \\ 
\hline \hline
$\mathrm{HWHM (pc) }$ \\
\hline \hline
$\mathrm{stars-alone}$ & $\mathrm{stars-coupled}$ &  $\mathrm{gas-alone}$ & $\mathrm{gas-coupled}$ \\
  
\hline                                                                                                                                   
457.1   &       380.5   &       755.0   & 159.2 \\ 
\hline
\end{tabular}
\end{threeparttable}
\end{table}

\begin{figure*} 
\centering
\includegraphics[height=2.3in,width=3.2in]{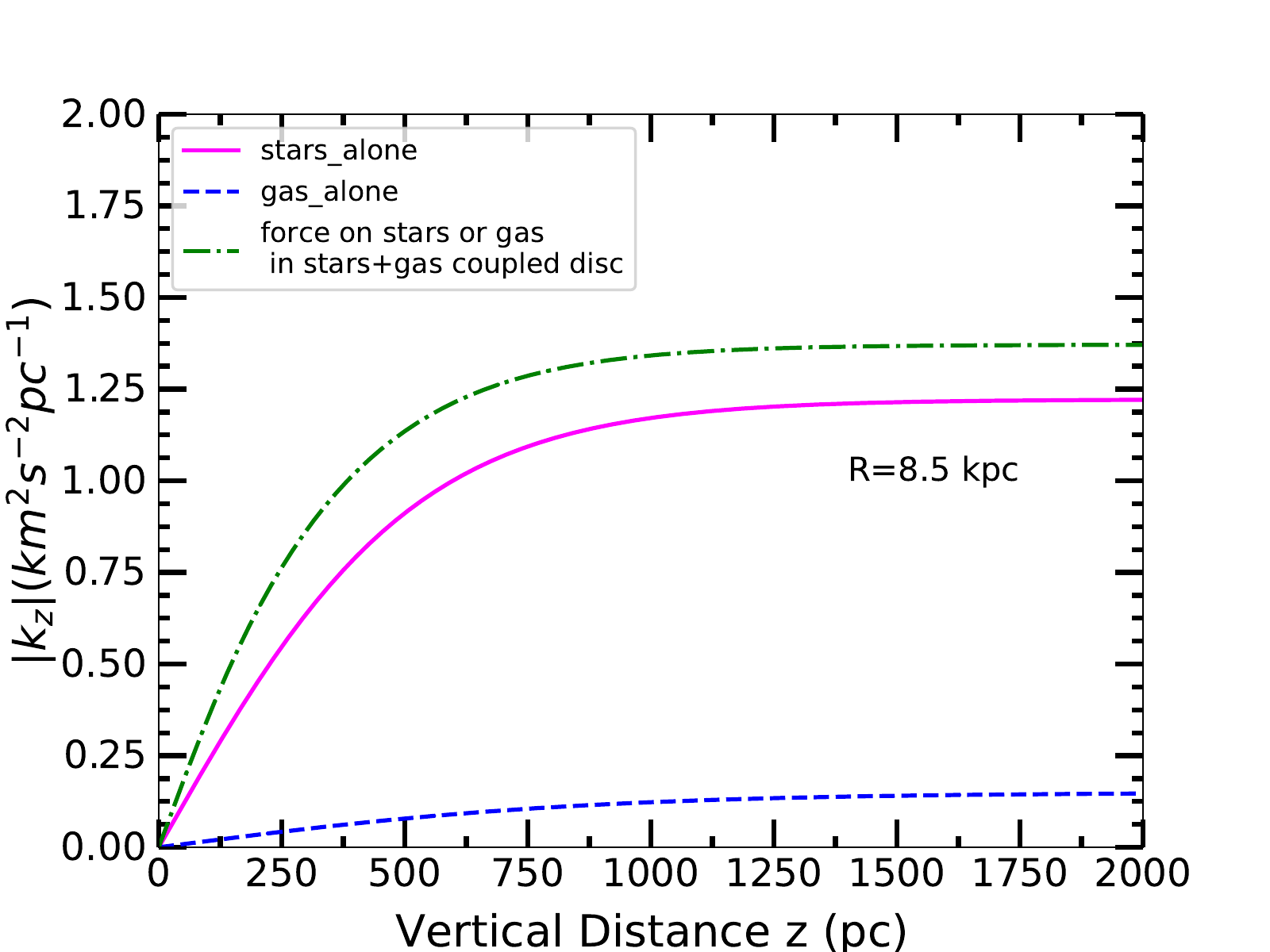}
\bigskip
\caption{Vertical force per unit mass, i.e. $|K_{z}|$ vs. $z$ at $R$= 8.5 kpc. The solid and dashed curves represent the force due to the self-gravity of stars and gas (HI), respectively. The dash-dotted curve represents the force arising from the gravitationally coupled stars-plus-gas system, which is higher than the forces due to the self-gravity of the individual components at each $z$.}
\label{chap4_label2}
\end{figure*}

We also calculated the self-gravitational forces of the stars-alone and gas-alone cases and compare them with the gravitational force of the coupled stars-plus-gas system in Fig.(\ref{chap4_label2}). The self-gravitational force for each single-component case was calculated using Eq.(\ref{chap4_eq:6}), as discussed in Section \ref{sec:formulate_chap4_single}. For the stars-plus-gas system, the coupled gravitational force is given by the right-hand side of Eq.(\ref{chap4_eq:14}). Each component is kept in hydrostatic equilibrium due to this coupled force. We calculated this force numerically from the left-hand side of this equation, substituting  $\rho_{i}$ and $\mathrm{d}\rho_{i}/dz$ for any component, both obtained numerically by solving Eq.(\ref{chap4_eq:16}). Figure \ref{chap4_label2} shows that at each $z$ distance from the mid-plane, this coupled force that keeps stellar and gas distributions in hydrostatic equilibrium in the coupled system is higher than the  one-component self-gravitational forces.

Motivated by these results, we now aim to compare the potential energy values of the stellar and gas distribution in the coupled case to the corresponding single-component cases.
We used Eq. (\ref{chap4_eq:2}) and the procedure outlined in Section \ref{sec:formulate_chap4_single} to calculate the potential energy per unit area of a stars-alone disk at the solar radius to be  $14257.7 \mathrm{M_{\odot}pc^{-2}km^{2}s^{-2}}$ ($3\times 10^{7} ergcm^{-2}$) and of a gas-alone disk to be 351.4 $\mathrm{M_{\odot}pc^{-2}km^{2}s^{-2}}$ ($7.4\times 10^{5} erg cm^{-2}$). The sum of the energies is then given by 14609.1$\mathrm{M_{\odot}pc^{-2}km^{2}s^{-2}}$. The $z$ ranges used in the integration of $\rho(z)~ \text{versus } z$ and in calculation of $K_{z}~\text{versus } z$ were chosen such that the energy values are numerically saturated (see Section \ref{sec:formulate_chap4_single}).

\begin{figure*} 
\centering
\includegraphics[height=2.3in,width=3.2in]{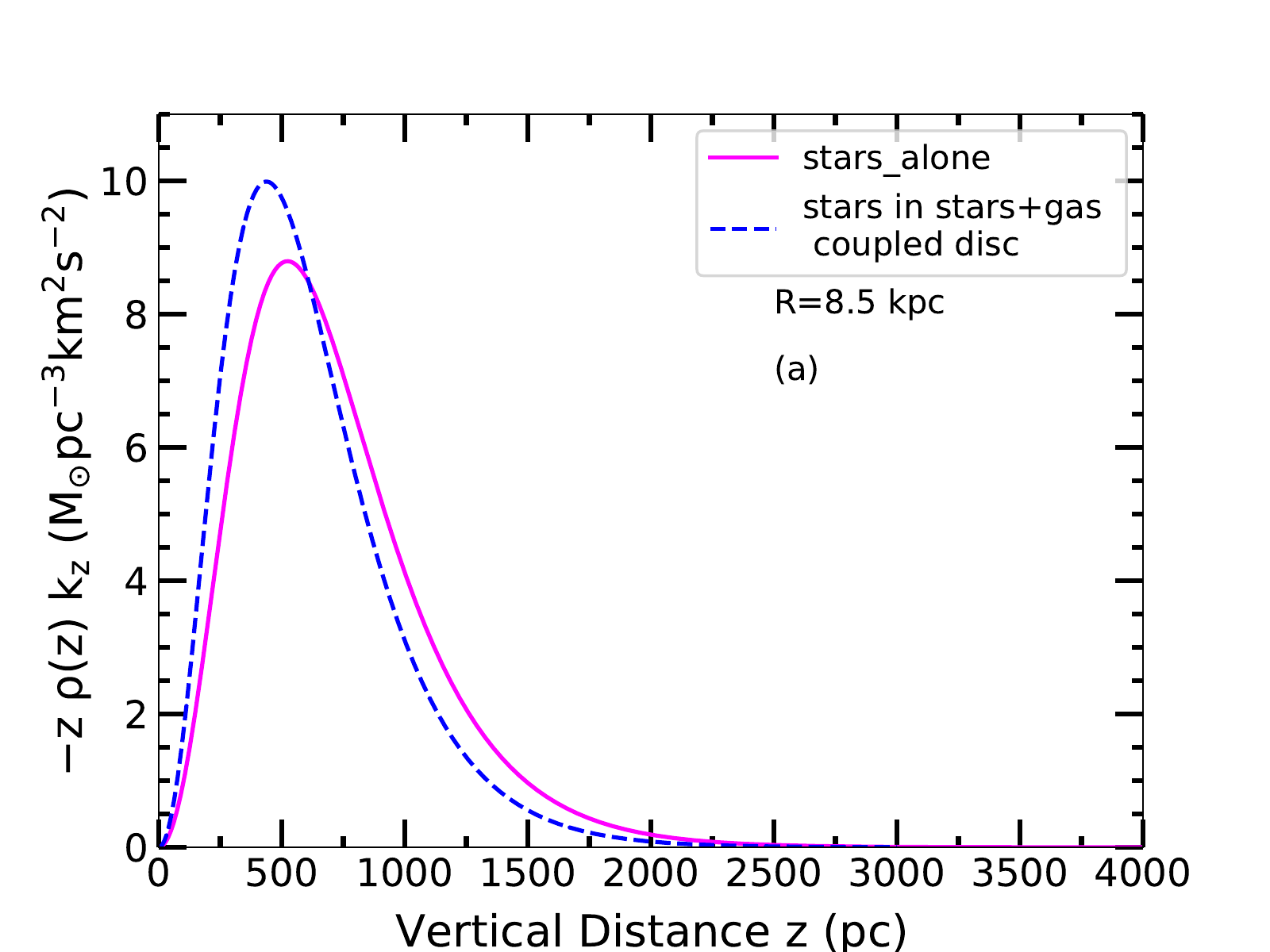}
\medskip
\includegraphics[height=2.3in,width=3.2in]{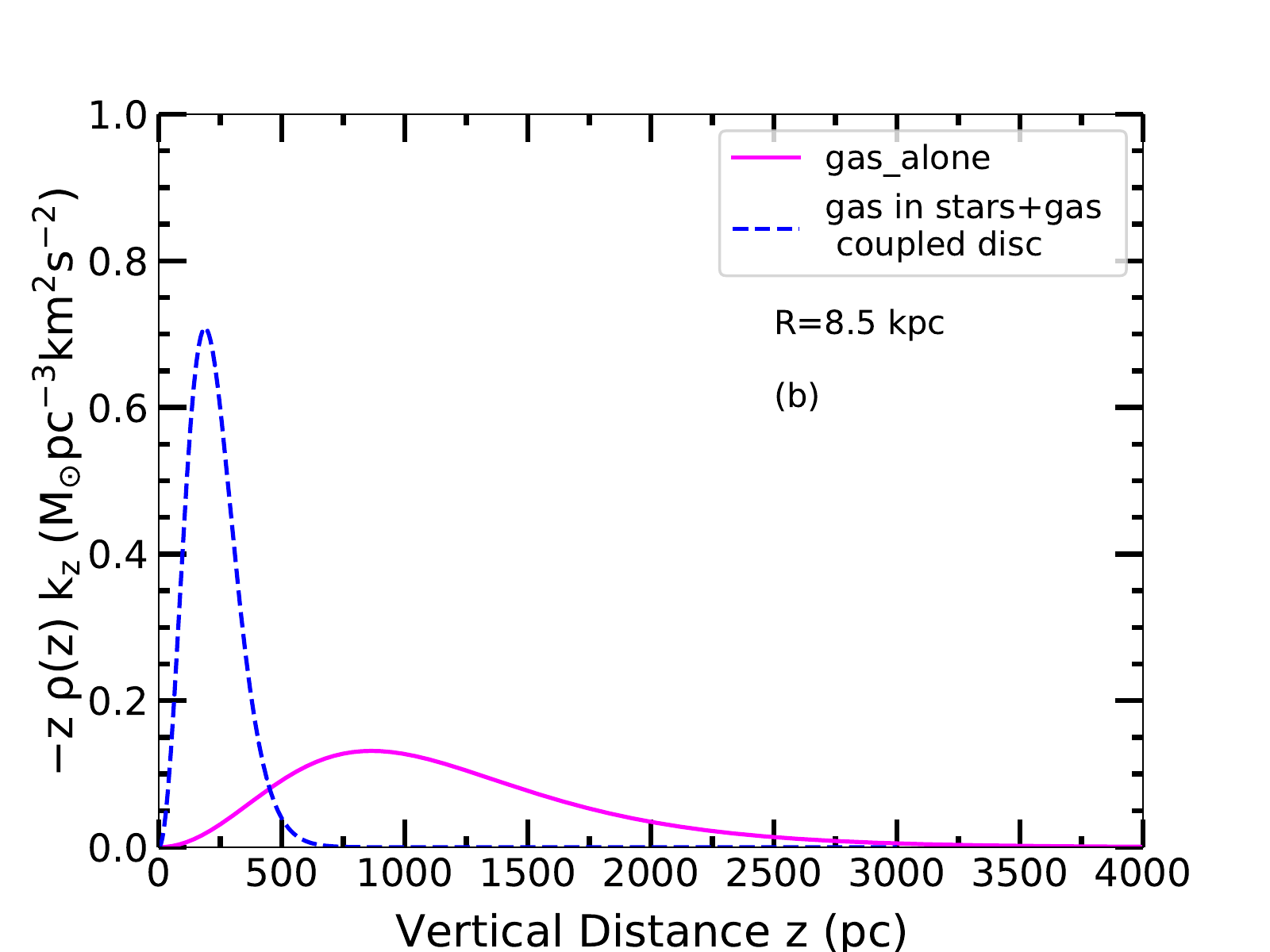}
\bigskip
\caption{Integrand $-z\rho(z)K_{z}$ vs. $z$ (from Eqs.(\ref{chap4_eq:2}), \ref{chap4_eq:13})) of $W$ (potential energy per unit area) for stars and gas (HI) at $R$=8.5 kpc. $\textit{Panel a}$: Energy integrand for stars in the stars-alone case (solid curve) and in the coupled stars-plus-gas case (dashed curve). Twice the area under the curves gives the energy per unit area of the stellar disk in the two cases, which is obtained to be the same.
$\textit{Panel b:}$ Energy integrand for gas (HI) in the gas-alone case (solid curve) and in the coupled stars-plus-gas case (dashed curve). Twice the area under the curves gives the energy per unit area of the gas disk in the two cases, which are obtained to be the same.}
\label{chap4_label3}
\end{figure*}

Now we calculate the potential energy per unit area of the coupled stars-plus-gas disk using Eq.(\ref{chap4_eq:13}) (using $\mathrm{K_{z,coupled}}$ and $\rho_{i}(z)$ obtained numerically as described in Section \ref{sec:chap4_distribution}) and find it to be 14609.02$\mathrm{M_{\odot}pc^{-2}km^{2}s^{-2}}$. This matches the sum of the energies for the stars-alone and the gas-alone cases within the numerical accuracy. Thus very interestingly, despite being in the gravitational force of the coupled system, the work done required to build up a stars-plus-gas disk turns out to be the same as the sum of the energies that would be required to build separate single-component self-gravitating stellar and gas disks. Importantly, the energy values for both stars and gas, in this case, obtained from Eq.(\ref{chap4_eq:13}) is the same as in the corresponding single-component self-gravitating cases within the numerical accuracy, that is, 14257.02 $\mathrm{M_{\odot}pc^{-2}km^{2}s^{-2}}$ and 352.0 $\mathrm{M_{\odot}pc^{-2}km^{2}s^{-2}}$ , respectively. We plot the energy integrand, that is, $-z\rho(z)K_{z}$ versus z in the single-component and in the coupled case for stars in Fig.(\ref{chap4_label3}a) and for gas in Fig.(\ref{chap4_label3}b). Twice the area under these curves gives the corresponding energy values. We note that the energy integrand is now redistributed along $z,$ conserving the area under the curve.

To investigate the physical reason that the same potential energy value was obtained, we simplified the expression of the energy further analytically. We discuss this in the following section.

\subsection{Analytical simplification of the expression of gravitational potential energy per unit area of a multi-component disk}
\label{sec:analytical}
First we considered a single-component case, for instance the stars-alone case. We substituted the expression of $K_{z}$ by the left-hand side of the hydrostatic balance equation (Eq.\ref{chap4_eq:3}) for a self-consistent distribution for a stars-alone disk into Eq.(\ref{chap4_eq:2}) and calculated it further as shown below,

\begin{align}
W_{\mathrm{stars-alone}} &= -2\int_{0}^{\infty} z K_{z,s} \rho_{s} \ dz  \nonumber \\
&= -2\int_{0}^{\infty} z \frac{\sigma^{2}_{z,s}}{\rho_{s}}\frac{d\rho_{s}}{dz} \rho_{s} \ dz.  \nonumber 
\end{align}
For an isothermal dispersion, we obtain this to be
\begin{equation}
W_{\mathrm{stars-alone}} = -2\sigma^{2}_{z,s}\int_{0}^{\infty} z \frac{d\rho_{s}}{dz} \ dz.  \nonumber 
\end{equation}
Applying the method of integration by parts, we obtain
\begin{equation}
W_{\mathrm{stars-alone}} = -2\sigma^{2}_{z,s} \left[ \bigg( z\rho_{s} \bigg)_{0}^{\infty} - \int_{0}^{\infty}\rho_{s} \ dz \right]  \label{chap4_eq:17}
.\end{equation}

\noindent Now at a very large $z$, theoretically, the density value is zero. Here, because $\rho_{s}$ falls off faster than $1/z$ (see Fig.\ref{chap4_label1}), the value of $z\rho_{s}$ at the upper limit will tend to zero. Using the numerically obtained solutions also, we can say that at the edge of the distribution, that is, at large $z$ by which the distribution is saturated and the value of the density is negligible, the product $z\rho_{s}$ becomes very small with respect to the other term. Thus the integration is 

\begin{align}
W_{\mathrm{stars-alone}} &\approx -2\sigma^{2}_{z,s}\left[ -\int_{0}^{\infty} \rho_{s} \ dz  \right]  \nonumber \\
&= -2\sigma^{2}_{z,s} \big( - \frac{\Sigma_{s}}{2} \big)  \nonumber \\
&= \sigma^{2}_{z,s}\Sigma_{s.} \label{chap4_eq:18}
\end{align} 

\noindent Thus the energy per unit area of the stellar disk is dependent only on the intrinsic parameters of the disk, namely its surface density and the vertical velocity dispersion. This expression is also valid for a gas-alone disk.  
Now we derive the corresponding analytical expression for the coupled stars-plus-gas disk in a similar way (using Eq.\ref{chap4_eq:13}), as given below,

\begin{equation}
W_{\mathrm{coupled}} = -2\int_{0}^{\infty}z K_{z,\mathrm{coupled}}~\rho_{s} \ dz -2\int_{0}^{\infty}z K_{z,\mathrm{coupled}}~\rho_{g} \ dz.   \nonumber
\end{equation}
Substituting $K_{z,\mathrm{coupled}}$ in terms of hydrostatic equilibrium of each component (Eq. \ref{chap4_eq:14}), we obtain
\begin{equation}
W_{\mathrm{coupled}} = -2\int_{0}^{\infty}\left[z \frac{\sigma^{2}_{z,s}}{\rho_{s}}\frac{d\rho_{s}}{dz}\rho_{s} \ dz\right] -2\int_{0}^{\infty}\left[z \frac{\sigma^{2}_{z,g}}{\rho_{g}}\frac{d\rho_{g}}{dz}\rho_{g} \ dz\right]. \nonumber
\end{equation}
\noindent Assuming isothermal dispersion, we obtain
\begin{align}
W_{\mathrm{coupled}} = -2\sigma^{2}_{z,s}\int_{0}^{\infty} \left[ z \frac{d\rho_{s}}{dz} \ dz\right] -2\sigma^{2}_{z,g}\int_{0}^{\infty} \left[ z \frac{d\rho_{g}}{dz} \ dz\right].   \label{chap4_eq:19}
\end{align}

\noindent This is similar to what is obtained for the single-component case. Applying integration by parts to each of the integrations and applying the physical argument at the large $z$ limit (as discussed above for the single component case), we obtain $W_{\mathrm{coupled}}$ to be
\begin{equation}
W_{\mathrm{coupled}}= \sigma^{2}_{z,s}\Sigma_{s} + \sigma^{2}_{z,g}\Sigma_{g} \label{chap4_eq:20}
.\end{equation}
\noindent This shows that the potential energy per unit area of each component (stars or gas) only depends on its intrinsic parameters, that is, the surface density and the vertical velocity dispersion, even within the coupled system. Consequently, the energy of any component remains the same in the coupled case as in the single-component case. After obtaining the general expression of energy of the multi-component system rigorously as in Eq.(\ref{chap4_eq:13}), we could simplify it analytically in a straightforward way, and thus could explain the constancy of the energy that was obtained numerically in Section \ref{sec:chap4_result_multi}.
In a similar way, the energy of any component in the three-component coupled system of stars and two gas components will also remain the same, as we show in Section \ref{sec:chap4_three}.

Based on these results, we can argue that physically, due to the joint gravity of the stars-plus-gas disk, the vertical distribution of stars and gas are now constrained toward the mid-plane. Thus due to the higher vertical force in the coupled case, the self-consistent distribution of each component is now effectively extended to a smaller vertical height, so as to conserve the energy per unit area. We also note that the joint gravity here works like an internal force within the system, and therefore it can just redistribute the energy within each of the two components without changing the total value of the energy.

However, instead of comparing the work done needed to build up the complete vertical mass distribution of the disk, we can compare the work done required to take only a unit, test mass from the $z=0$ plane to a certain finite height, discussed in the following section. This could  verify whether the disk is more likely to resist distortion for the constrained distribution resulting in a coupled case.

\subsection{Work done to raise a unit test mass from the mid-plane to a finite height}
\label{sec:chap4_unit_mass}

\begin{figure*} 
\centering
\includegraphics[height=2.3in,width=3.2in]{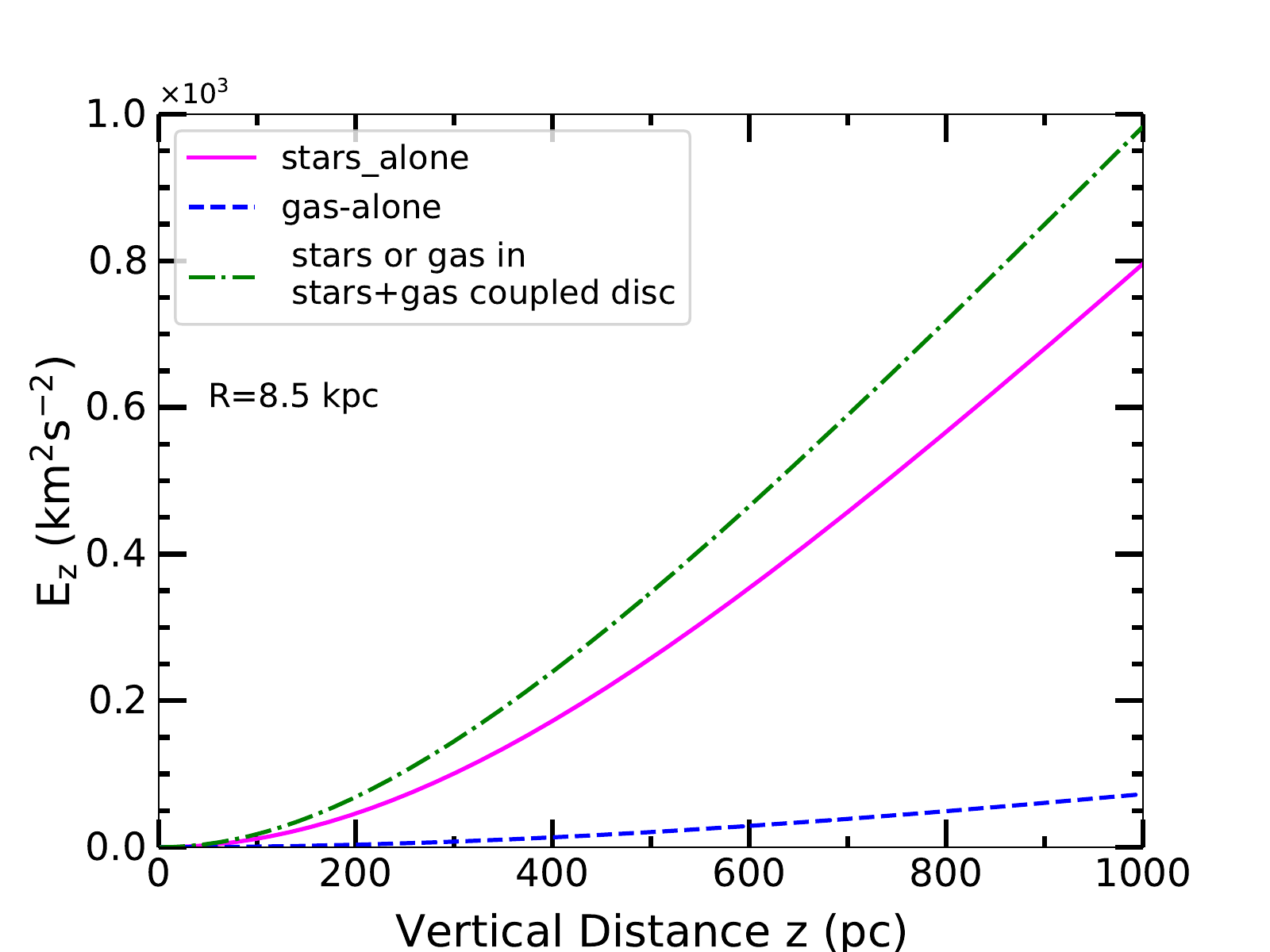}
\bigskip
\caption{Work done ($\mathrm{E_{z}}$) to raise a unit test mass from the mid-plane to any vertical height as a function of $z$ (shown upto z=1 kpc here) at $R$=8.5 kpc (solar neighbourhood) for the stars-alone case against its self-gravity (solid curve), for the gas-alone case against its self-gravity (dashed curve), and for stars or gas in the stars-plus-gas coupled system against their coupled gravity (dash-dotted curve). The work done required to raise a unit mass at any vertical height in the last case is highest. This shows that stars or gas in the coupled system are more strongly bound to the mid-plane than in the corresponding single-component cases.}
\label{chap4_label4}
\end{figure*}
\label{sec:chap4_result_unit}
We derived the expression of the work done or the energy required to raise a unit test mass from the mid-plane to a certain height $h$ in a single-component (stars or gas) disk and in the coupled two-component stars-plus-gas disk. The work has to be done against the self-gravity of stars (gas) and the joint gravity of stars plus gas disk, respectively. We note that this is precisely the measure of the gravitational potential at any height (\citealt{1984a_Bahcall,1984b_Bahcall}) in these cases.

For a single-component disk (of stars or gas), this work done is given as 

\begin{equation}
E_{z,i}=-\int_{0}^{h} K_{z,i} \ dz  \label{chap4_eq:21}
,\end{equation}

\noindent where $K_{z,i}$ represents the self-gravity of the disk.

\noindent In the gravitationally coupled disk of stars plus gas, both stellar and gas distribution are subject to the same coupled force. Therefore, the work done on a unit mass of the stellar or the gas distribution is

\begin{equation}
E_{z,i,\mathrm{coupled}}=-\int_{0}^{h} K_{z,\mathrm{coupled}} \ dz  \label{chap4_eq:22}
.\end{equation}

\noindent We note that in each case the energy is positive. Due to a higher vertical force at each z, as was shown in Fig.(\ref{chap4_label2}), the work done to take the unit mass of the stellar (gas) disk to the same height in the coupled case will be higher than in the corresponding single-component case. We show the work done in these cases as a function of vertical height in Fig.(\ref{chap4_label4}).

This shows that to raise a unit mass from the mid-plane to a certain vertical distance, more work is required in the coupled case than that for a single-component case. Thus stars are more strongly bound to the mid-plane of the Galaxy in the coupled case than in the single-component case, and thus the stellar disk will be able to offer more resistance to a given external tidal encounter. In this case, the stellar disk is therefore less likely to be thickened (\citealt{1996_Walker}). Moreover, due to the constrained distribution, the stellar mass distribution is more concentrated towards the mid-plane.  This increases the effective gravity of the stellar disk near the mid-plane and helps it to resist external perturbations, which could have led to the generation of warps \citep{Pranav_Jog_2010}. Thus the constraining effect of gas on stars makes the stellar disk less likely to be disturbed. A detailed N-body simulation will be able to show this clearly. This is beyond the scope of this paper. 

Similarly, a unit mass of the gas disk constrained by the stellar gravity is more strongly bound in the coupled system than in the gas-alone case, and hence significantly more work is required to raise a unit mass of the gas disk to a certain height than in the gas-alone case. Thus gas disk in the coupled system is less likely to be disturbed than the gas-alone case. Furthermore, we note that far more work is required to raise gas to a certain height in the coupled case compared to gas-alone case than the corresponding work required for stars because stars are the more massive component and have a stronger effect on gas. 

\subsection{Calculation of the potential energy of a three-component disk}
\label{sec:chap4_three}
For the sake of completeness, we next studied the gravitational potential energy per unit area of a gravitationally coupled three-component galactic disk, consisting of stars and two gas components. The three components were taken to be coplanar with the same mid-plane at z=0. For illustration, we added $\mathrm{H_{2}}$ as the second gas component in addition to HI, as seen in the inner Galaxy. We note that this three-component treatment is essential in the inner Galaxy. 

For the three-component system of stars, HI, and $\mathrm{H_{2}}$, the gravitational potential energy per unit area of the disk is the work done to build a column of gravitationally coupled stars, HI, $\mathrm{H_{2}}$, of unit cross-section, together from $z=0$. The Poisson equation for this system is given as 

\begin{equation}
\frac{d^{2}\Phi_{\mathrm{coupled}}}{dz^{2}}  = 4\pi G (\rho_{s}+\rho_{\mathrm{HI}}+\rho_{\mathrm{H_{2}}})\label{chap4_eq:23}
.\end{equation}

\noindent Following the same procedure as in Section {\ref{sec:chap4_multi}}, we derived the potential energy per unit area of the three-component disk as 

\begin{equation}
W_{\mathrm{coupled}} = \int_{-\infty}^{\infty}z~ \frac{d\Phi_{\mathrm{coupled}}}{dz}(\rho_{s}+\rho_{\mathrm{HI}}+\rho_{\mathrm{H_{2}}})  \ dz  \label{chap4_eq:24}
,\end{equation}

\noindent which can be further expressed as 
\begin{align}
W_{\mathrm{coupled}} &= -2\int_{0}^{\infty}z K_{z,\mathrm{coupled}} ~\rho_{s} \ dz - 2\int_{0}^{\infty}zK_{z,\mathrm{coupled}} ~\rho_{\mathrm{HI}} \ dz \nonumber \\ 
&- 2\int_{0}^{\infty}zK_{z,\mathrm{coupled}} ~\rho_{\mathrm{H_{2}}} \ dz. \label{chap4_eq:25}
\end{align}

\noindent These three integrations can be considered to represent the potential energy per unit area of the stellar disk, HI disk, and $\mathrm{H_{2}}$ disk in the three-component coupled system. 

Now for a three-component system, the hydrostatic balance of each component is determined by the joint gravitational force from stars, HI, and  $\mathrm{H_{2}}$, and is given by 

\begin{equation}
\frac{\sigma^{2}_{z,i}}{\rho_{i}}\frac{d\rho_{i}}{dz}=K_{z,s}+K_{z,\mathrm{HI}}+K_{z,\mathrm{H_{2}}} \label{chap4_eq:26}
,\end{equation}
\noindent where $i$ represents stars $(s)$ or HI or $\mathrm{H_{2}}$, and the right-hand side of the equation represents the vertical force of the coupled system. Combining this equation with Eq.(\ref{chap4_eq:23}), we write the joint hydrostatic balance-Poisson equation for the three-component system as 

\begin{equation}
\frac{\mathrm{d}^{2}\rho_{i}}{\mathrm{d}z^{2}} = \frac{\rho_{i}}{\sigma^{2}_{z,i}}\left[-4\pi G\left(\rho_{\mathrm{s}}+ \rho_{\mathrm{HI}}+ \rho_{\mathrm{H_{2}}}\right)\right] +\frac{1}{\rho_{i}}\left(\frac{\mathrm{d}\rho_{i}}{\mathrm{d}z}\right)^{2}  \label{chap4_eq:27}
.\end{equation}

\noindent These coupled equations were solved to obtain $\rho_{i}~\text{versus } z$ for stars, HI, and $\mathrm{H_{2}}$ following the same method as discussed in Section \ref{sec:chap4_distribution}.
Here, we discuss the results for the stellar distribution alone for simplicity because the results for the gas components follow a similar trend as for stars, as seen for the two-component case in Section \ref{sec:chap4_result_multi} and \ref{sec:chap4_unit_mass}. We chose $R$=4.5 kpc to illustrate the results here. The stellar surface density at $R$=4.5 kpc was calculated to be 157.06 $\mathrm{M_{\odot}pc^{-2}}$ \citep{1998_Mera}. The vertical velocity dispersion of stars was obtained to be  28.2 $km s^{-1}$ (\citealt{1989_Lewis_Freeman}) and assuming the vertical-to-radial dispersion ratio to be the same as in the solar neighbourhood, see Section \ref{sec:chap4_input}. We chose this radius to prominently show the effect of the third component, namely $\mathrm{H_{2}}$, on stars. The surface density of $\mathrm{H_{2}}$ is 19.7 $\mathrm{M_{\odot}pc^{-2}}$ at this radius, which is significantly higher than that of HI, which is 4.6 $\mathrm{M_{\odot}pc^{-2}}$ (\citealt{1987_Scoville}). The vertical velocity dispersion of HI is 8 $\mathrm{kms^{-1}}$, as it was in the two-component case (Section \ref{sec:chap4_input}), and that of $\mathrm{H_{2}}$ is 5 $\mathrm{kms^{-1}}$ at this radius (\citealt{1987_Scoville}).

\begin{figure*} 
\centering
\includegraphics[height=2.3in,width=3.2in]{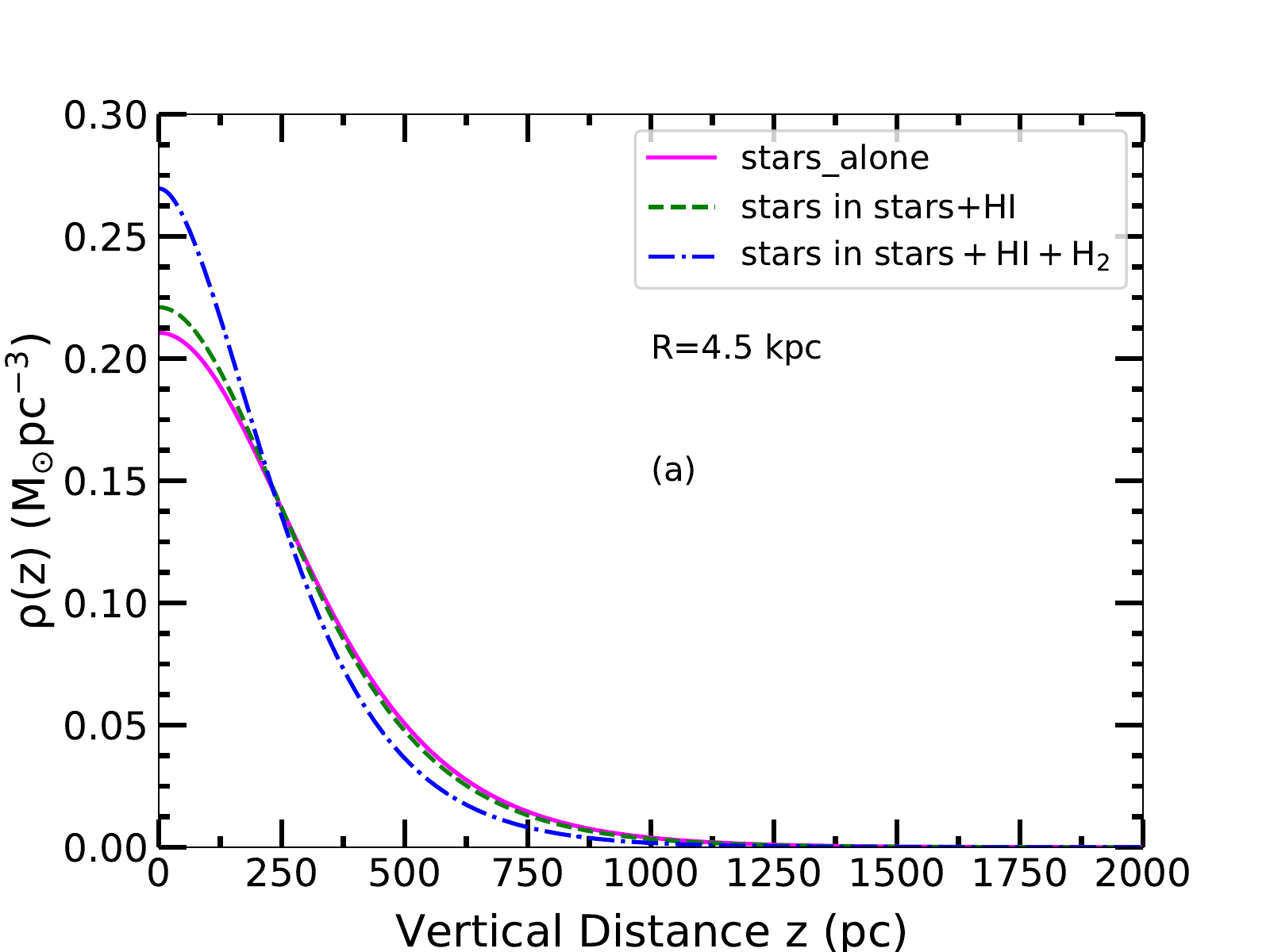}
\medskip
\includegraphics[height=2.3in,width=3.2in]{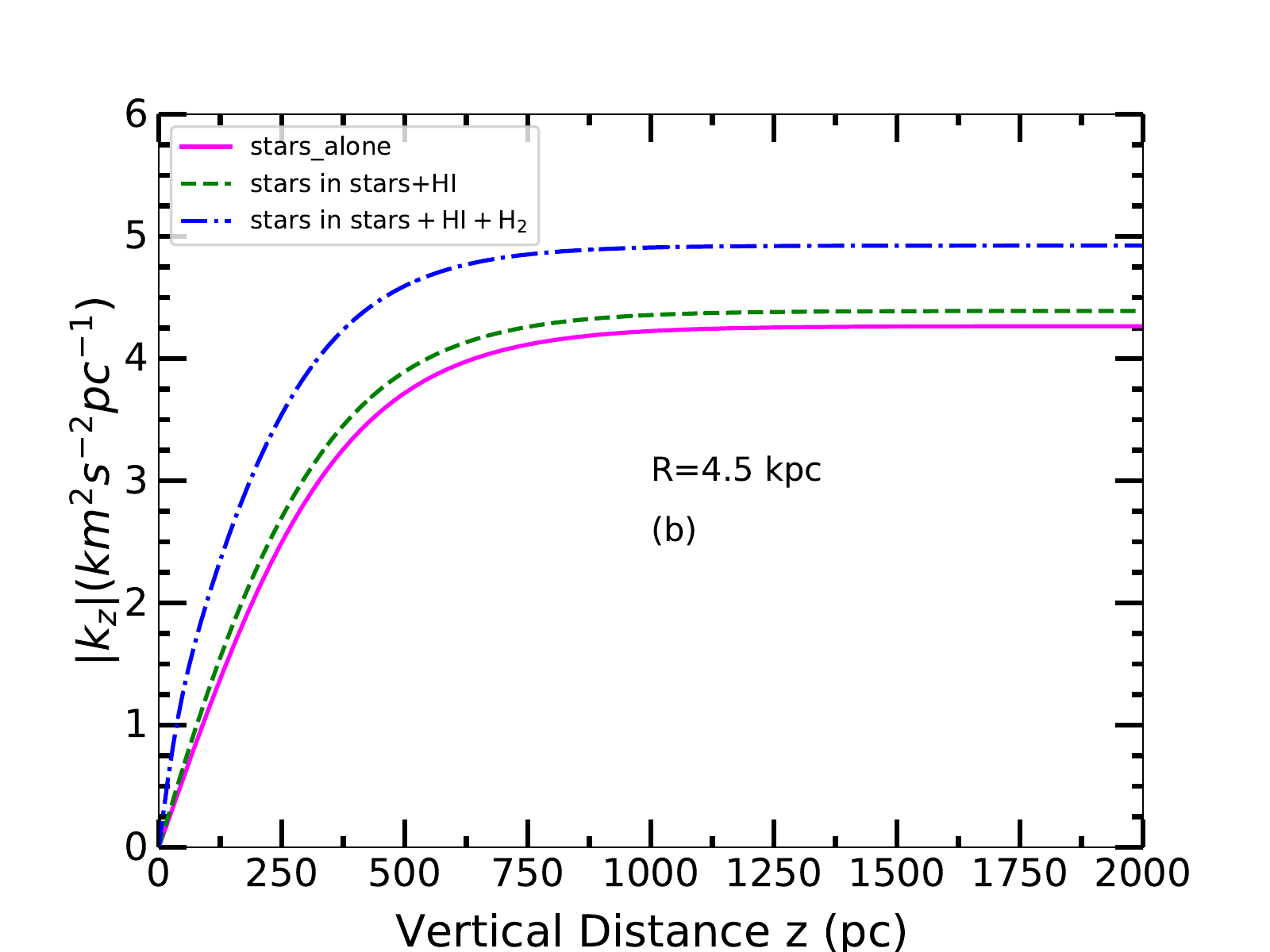}
\medskip
\includegraphics[height=2.3in,width=3.2in]{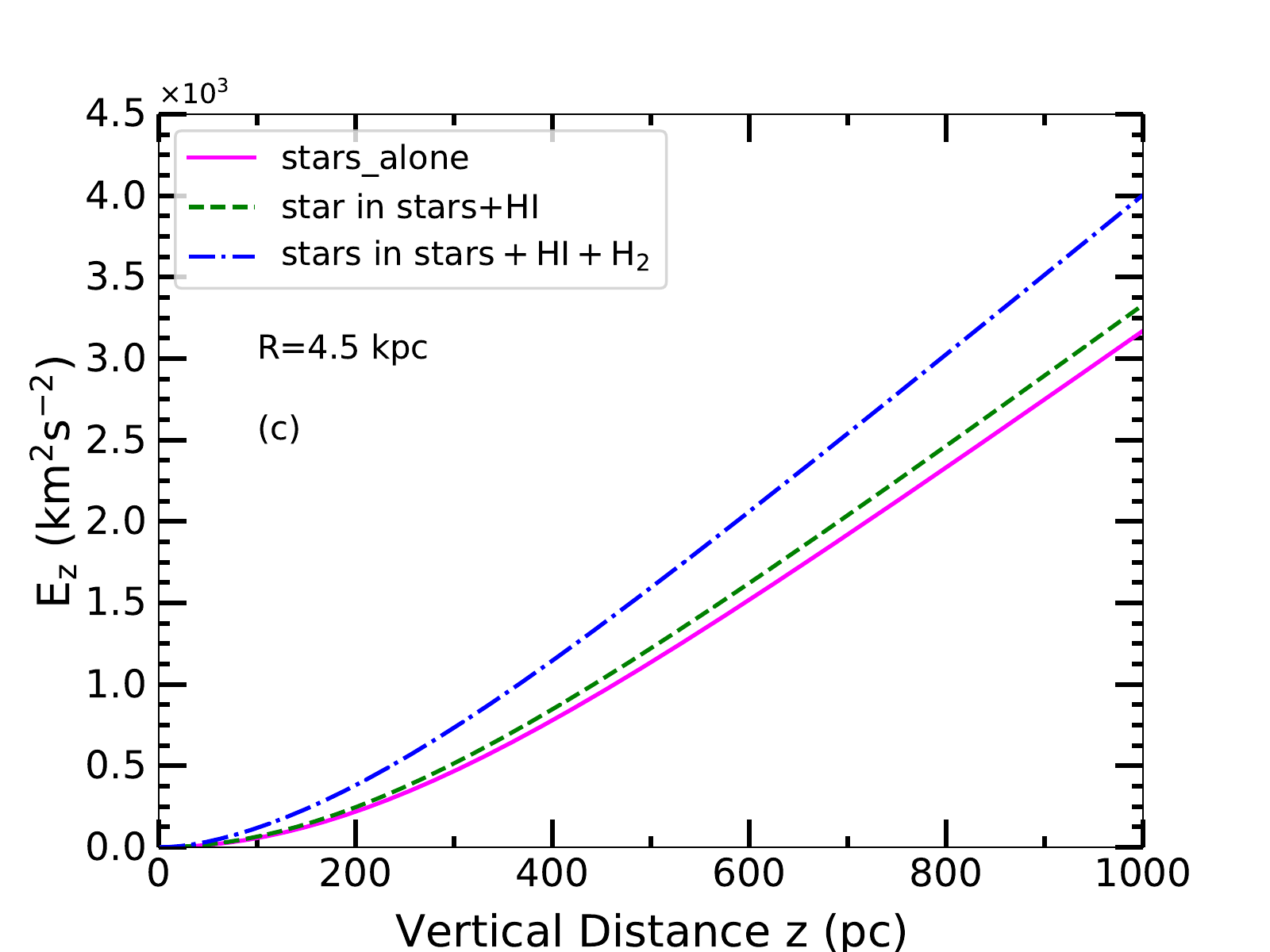}
\caption{Results for stellar distribution in the three-component, gravitationally coupled stars plus HI plus $\mathrm{H_{2}}$ disk, two-component stars plus HI disk, and stars-alone case at $R$=4.5 kpc. $\textit{\textbf{Panel a}}$: Stellar vertical density distribution, $\rho(z)$ vs. $z$ is shown for stars-alone case (solid curve), gravitationally coupled stars plus HI case (dashed curve), and stars plus HI plus $\mathrm{H_{2}}$ case (dash-dotted curve). The gas gravity constrains the distribution of stars towards the mid-plane by raising its mid-plane density and reducing its scale height value. This constraining effect is highest in the three-component case. At the same time, the relative contribution of $\mathrm{H_{2}}$ is higher than that of HI as the surface density of $\mathrm{H_{2}}$ is higher than HI at $R$=4.5 kpc. $\textit{\textbf{Panel b}}$: Vertical force per unit mass, i.e. $|K_{z}|$ vs. $z$, acting on stars is shown due to the self-gravity of stars (solid curve), due to the force from the gravitationally coupled stars plus HI disk (dashed curve), and due to the force from the coupled stars plus HI plus $\mathrm{H_{2}}$ disk (dash-dotted curve). The force at any $z$ is highest in the three-component case. $\textit{\textbf{Panel c}}$: Work done ($\mathrm{E_{z}}$) to raise a unit test mass from the mid-plane to any vertical height as a function of $z$ (shown up to z=1 kpc here) for stars against its self-gravity (solid curve), against the gravitational force from the coupled stars plus HI disk (dashed curve), and against the gravitational force from the coupled stars plus HI plus $\mathrm{H_{2}}$ disk (dash-dotted curve). The work done in the three-component case is highest. This shows that stellar distribution in a three-component system is more strongly bound to the mid-plane than in a two-component system.}
\label{chap4_fig5}
\end{figure*}

We show the results for stellar distribution in Fig. \ref{chap4_fig5} in the three-component system, in the two-component system (stars plus HI), and in the stars-alone case. Fig. \ref{chap4_fig5}a shows that the addition of a second gas component ($\mathrm{H_{2}}$ here) constrains the stellar distribution towards the mid-plane by raising the mid-plane density value and reducing the disk thickness value compared to the values in the two-component case. We note that due to a higher surface density of $\mathrm{H_{2}}$, the constraining effect due to this component on stars is more prominent than that due to HI. Fig. \ref{chap4_fig5}b shows that the coupled gravitational force per unit mass (calculated numerically using Eq.\ref{chap4_eq:26}) that keeps the stellar distribution in the hydrostatic equilibrium in the three-component system is higher than that of the two-component system and the stars-alone case at all $z$. 

Despite the higher constraining effect in the three-component system, the potential energy per unit area of the stellar distribution, calculated numerically from Eq.(\ref{chap4_eq:25}), is found to be 124900.4 $\mathrm{M_{\odot}pc^{-2}km^{2}s^{-2}}$ ($2.6\times 10^{8} erg ~cm^{-2}$) , which is the same as that found for stars in the two-component system and in the stars-alone system at $R$=4.5kpc. This result is expected because we find the analytical expression of the total energy per unit area of the three-component coupled disk to be $\mathrm{W_{coupled}}=\sigma^{2}_{z,s}\Sigma_{s} + \sigma^{2}_{z,\mathrm{HI}}\Sigma_{\mathrm{HI}} + \sigma^{2}_{z,\mathrm{H_{2}}}\Sigma_{\mathrm{H_{2}}}$, derived following a similar method as discussed for the two-component system in Section \ref{sec:analytical}. Thus, we note that the energy of each component remains unchanged.

However, when we calculate the work required to take a unit test mass of stellar distribution from the mid-plane to a certain height $h$ in the above three cases following the method discussed in Section \ref{sec:chap4_unit_mass}, we find that the work done is highest in the three-component case at any $z$. We show the work done as a function of $z$ corresponding to the three cases in Fig. \ref{chap4_fig5}c. This shows that stars are more strongly bound to the mid-plane of the Galaxy in the three-component case than in the two-component system, and thus the stellar disk is able to offer more resistance to a given external tidal encounter. Thus the constraining effect of the two gas components will make the stellar disk less likely to be disturbed due to external perturbations. Although we do not show results for gas here for conciseness, the stellar component, being more massive, has a higher effect on making the gas disk less likely to be disturbed due to perturbations than the gas-alone case (as was already seen for the two-component case in Sections \ref{sec:chap4_result_multi} and \ref{sec:chap4_unit_mass}).

\section{Discussion}
\label{sec:chap4_discussions}
We discuss a few general implications of the model developed in this paper below.
First, the main aim of this paper was to determine how the gravitational potential energy per unit area of the disk components changes in view of the constraining effect in the coupled case. The energy of the components turned out to be the same as in their single-component self-gravitating cases, which is in contrast to our initial expectation. Therefore, the question is whether stars and gas are more strongly bound in the coupled case. Our results have shown that it requires a higher amount of energy to raise the unit mass of a component to a certain vertical height in the multi-component case than in its single component case, even though the component contains the same potential energy per unit area in both cases. Thus each component is more strongly bound to the mid-plane in the multi-component system. 
\\\\
\noindent Second, we note that the potential energy per unit area of any disk component is dependent on its surface mass density and its vertical velocity dispersion. This implies that the magnitude of the energy depends on the component chosen at a given galactocentric radius and also on the galactocentric radius for a given component, as the above parameters vary along radius. We note that at the solar radius as well as at $R$=4.5 kpc, the surface density and the vertical velocity dispersion of stars are so much higher than gas that the energy value per unit area for the stellar disk is much higher than gas. The stellar disk is more extended vertically than the gas disk, and hence we would expect it to be more disturbed by a given external tidal encounter, for instance from a passing satellite galaxy. However, the stellar disk has a higher potential energy per unit area. Although it is more extended, it is therefore less likely to show the effect of a given tidal disturbance. This trend can be confirmed by numerical simulations of an encounter. This is beyond the scope of this paper.
\par
We also note that we have considered only gravitational interaction for both stellar and gas disks. We did not consider any gas dynamical phenomenon.
\\\\
\noindent Third, in the above cases, we have taken two (three) components with different dispersions to identify them as stars and one (two) gas components. It may be an interesting physical question to ask what happens when the stellar disk is divided artificially into n number of components. Following a similar analysis, we find again that the total energy of the components remain unchanged, as is the case for the components with different dispersions. Interestingly, in this case, the net distribution is not vertically more constrained, as expected physically. For example, following the numerical analysis in Section \ref{sec:chap4_distribution}, we checked that if we were to divide the disk of $\Sigma$ into two sub-components of $\Sigma_{1}$ and $\Sigma_{2}$ with the same dispersions, the net distribution in the coupled case would be identical to the one-component case with $\Sigma$. 
\\\\
\noindent Fourth, the treatment given in the paper is general. Although we have applied it to stars and gas case, it can be applied to n number of stellar sub-components in the Galactic disk as well. Such sub-components have been identified from recent observed data, for example from Gaia (\citet{bovy_2017}, \citet{2018_Hagen_Helmi} etc).  If the surface density dispersions are known for these components, then the potential energy per unit area of these components can be determined quantitatively. 

\section{Conclusions}
\label{sec:chap4_conclusions}

It has been shown earlier that in a multi-component gravitationally coupled stars-plus-gas disk, the self-consistent vertical distribution of stars is constrained closer to the mid-plane (\citealt{2018_Sarkar}). In order to understand the implications of this for the energetics of the disk, we obtained the potential energy per unit area at a given galactocentric radius for a multi-component galactic disk. This was obtained as an integration over the vertical density ($\rho(z)$), the gravitational force, and vertical distance $z$. To do this, we followed the method developed by \citet{1967_Camm} for a single-component self-gravitating disk and explicitly derived the corresponding expression for the multi-component case. For a self-consistent distribution we obtained the density distribution and force as a function of $z$ by numerically solving the joint hydrostatic balance and Poisson equation for the coupled case.

\noindent 
1. We find that the net gravitational potential energy for the stars and gas remain unchanged to that in the single-component cases. This is a surprising result and can be understood by simplifying analytically the general expression for the potential energy that we obtain for the multi-component system. We noted that the potential energy per unit area of each component depends only on its intrinsic parameters, that is, the vertical velocity dispersion and the surface density, in the single-component as well as in the multi-component case. 

Physically, the energy values remain unchanged because due to the higher joint gravity in the coupled case, the distribution of each component is constrained closer to the mid-plane and thus has a less effective vertical thickness to conserve the energy per unit area. We note that the joint gravity works here like an internal force within the system, and therefore it can just redistribute the energy within each component itself without changing the total value of energy for each component.

\noindent 
2. However, due to the constrained distribution in the coupled cases, the work required to raise a unit test mass to a given height is higher than that in the single-component case. Thus, while constraining in a coupled case does not correspond to any additional gravitational energy in the system, it does indicate that each component in a disk in the coupled case is more strongly bound to the mid-plane. Furthermore, the stellar disk has a higher potential energy per unit area than the gas disk. For a given tidal encounter, the stellar disk is therefore less likely to be disturbed than a gas disk, even though it is more vertically extended.

\medskip
\begin{acknowledgements}
We thank the anonymous referee for very useful comments. We would like to thank Tarun Deep Saini for discussions which helped in clarifying our ideas for developing the problem. S.S. thanks CSIR for a fellowship, and C.J. thanks the DST for support via J.C. Bose fellowship (SB/S2/JCB-31/2014).
\end{acknowledgements}

\bibliographystyle{aa}
\bibliography{ms}
\end{document}